\documentclass[12pt]{article}

\usepackage{graphicx}

\newcommand{\beq}{\begin{equation}}
\newcommand{\eeq}{\end{equation}}
\newcommand{\bea}{\begin{eqnarray}}
\newcommand{\eea}{\end{eqnarray}}
\newcommand{\br}{{\bf r}}

\begin{document}
\pagestyle{empty}

\begin{center}
{\LARGE Analysis and Modelling of Subthreshold Neural Multi-electrode Array Data by Statistical Field Theory}

\vspace*{5mm}
{\large M{\aa}ns Henningson}

\vspace*{3mm}
{Department of Physics, Chalmers University of Technology}\\
{S-412 96 G\"oteborg, Sweden}\\
{\tt mans@chalmers.se}

\vspace*{5mm}
{\large Sebastian Illes}

\vspace*{3mm}
{Institute of Neuroscience and Physiology, University of Gothenburg}\\
{Box 430, S-405 30 G\"oteborg, Sweden}\\
{\tt sebastian.illes@neuro.gu.se}

\vspace*{10mm}
{\large Abstract}:
\end{center}
\noindent
Multi-electrode arrays (MEA) are increasingly used to investigate spontaneous neuronal network activity. The recorded signals comprise several distinct components: Apart from artefacts without biological significance, one can distinguish between spikes (action potentials) and subthreshold fluctuations (local fields potentials). Here we aim to develop a theoretical model that allows for a compact and robust characterization of subthreshold fluctuations in terms of a Gaussian statistical field theory in two spatial and one temporal dimension. What is usually referred to as the driving noise in the context of statistical physics is here interpreted as a representation of the neural activity. Spatial and temporal correlations of this activity give valuable information about the connectivity in the neural tissue. We apply our methods on a dataset obtained from MEA-measurements in an acute hippocampal brain slice from a rat. Our main finding is that the empirical correlation functions indeed obey the logarithmic behaviour that is a general feature of theoretical models of this kind. We also find a clear correlation between the activity and the occurence of spikes. Another important insight is the importance of correcly separating out certain artefacts from the data before proceeding with the analysis.

\newpage
\pagestyle{plain}

\section{Introduction}
The multi-electrode array system (MEA) is becoming an increasingly important tool for investigations of neural activity, both in ex vivo brain tissue (e.g. a hippocampal slice preparation from rat or mouse \cite{Egert-Heck-Aertsen}) and in in vitro neuronal cultures (e.g. from embryonic rodent brain tissue \cite{Illes-etal} or human stem cells \cite{Heikkila-etal}). This technology permits simultaneous long-term recordings from a fairly large number of extra-cellular electrodes. See e.g. \cite{Nam-Wheeler}\cite{Spira-Hay} for general reviews of multi-electrode array technology. 

Each electrode records alterations of the field potential elicited by spike activity of one or a few neurons in close vicinity of it. Extracellular spikes have an amplitude of 10-500 $\mu V$, and are considered as a manifestation of the intracellular action potential which has a much higher amplitude of 100 mV. Many methods have been developed for the detection and sorting of spike events (see e.g. \cite{Cotterill-Charlesworth-Thomas-Paulsen-Eglen} for a recent review), and analysis of the statistical properties of spike trains is one of the major modes of investigating neural activity (see e.g. \cite{Rieke-Warland-deRuyter-Bialek} for a pedagogical introduction to this field). 

The focus of the present paper will however not be on the spikes, but rather on the subthreshold behaviour of the potential during interspike intervals. This is often referred to as the `local field potential' (LFP). The extraction of these subthreshold fluctuations out of MEA-data streams which are Òcontaminated Ò with artifacts and spiking activity (see e. g. \cite{Waldert-Lemon-Kraskov}), as well as to develop the appropriate mathematical approaches applicable to describe their properties, are challenging issues in the research field of neuroscience. 

The local field potential is usually assumed to be a superposition of contributions from the neural activity in a fairly large neighborhood of an electrode and also gets modified by other types of cells than neurons. Typically it has an amplitude of a few $\mu V$, i.e. much less than the spikes. In contrast to the rather stereotyped spike waveforms from individual neurons, the local field potential has a much more stochastic appearance, reflecting its origins from a large and rather heterogenous ensemble of neurons \cite{Illes-Jakab}. Furthermore, spikes can only be recorded in relatively close vicinity to the recording electrode while the distance between recording electrode and local field potentials source can be several micrometers and millimeters \cite{Destexhe-Bedard}. Thereby, the local field potential recorded in isolated brain structures actually carries important information about the transport properties of the intra-cellular medium \cite{Bedard-Destexhe}, the spatial and temporal structure of the neural activity \cite{Linden-Tetzlaff}\cite{Destexhe-Bedard}, and also the connectivity of neurons within brain tissues \cite{Reichinnek-Kunsting}. 
There are different functional aspects of neuronal circuits which can be revealed by modelling and analyzing local field potentials. Current-source density analysis is used to reveal the neural source of recorded local field potentials \cite{Ness-Chintaluri} which is still controversial \cite{Riera-Ogawa}. Since the pioneering work of Berger in the 1920s, local field potential band-separations techniques has been devoted to analyzing the local field potential in the frequency domain (see e.g. \cite{Buzsaki}) with the aim of identifying physiologically and pathophysiologically relevant frequency bands. In addition, decomposing local field potentials into different frequency bands are used to correlate them to cognitive or motoric function as well as neuronal spiking activity. The aim here is to decipher brain activity in controlling perception, cognitive, motoric function and, in particular for the hippocampus,  memory and learning abilities. Thus, the analysis of spike-LFP relationship represents another approach. Huge efforts are currently being done by model inversion approaches by creating Òartificial neuronal networksÓ which produce realistic local field potentials. In this approach, models of neural networks are combined with experimental data to identify the best-fit model. However, studies in which the applicability of mathematical or physical theories is evaluated by comparing the result of the model with experimental data are still rare but needed \cite{Ness-Chintaluri}. 

We aim to describe the spatiotemporal properties of subthreshold fluctuations in the rat hippocampal circuit by applying a mathematical description based on Gaussian statistical field theory to MEA data.
Our study puts more emphasis on the spatio-temporal structure of correlations and less on oscillations. A generic feature of theoretical models in two spatial and one temporal dimension is a logarithmic behaviour at short scales; as we will see this prediction is convincingly confirmed by our empirical material and is in a sense our main conceptual finding. From a perspective of practical electrophysiology, we would also like to emphasize the importance of correctly separating out the effects of certain artefacts without biological relevance before proceeding with the analysis.

\section{Methods}
\subsection{Multi-electrode array setup}
\begin{figure}[h!] 
\centering
\includegraphics[width = \linewidth, trim = 0mm 140mm 0mm 0mm, clip]{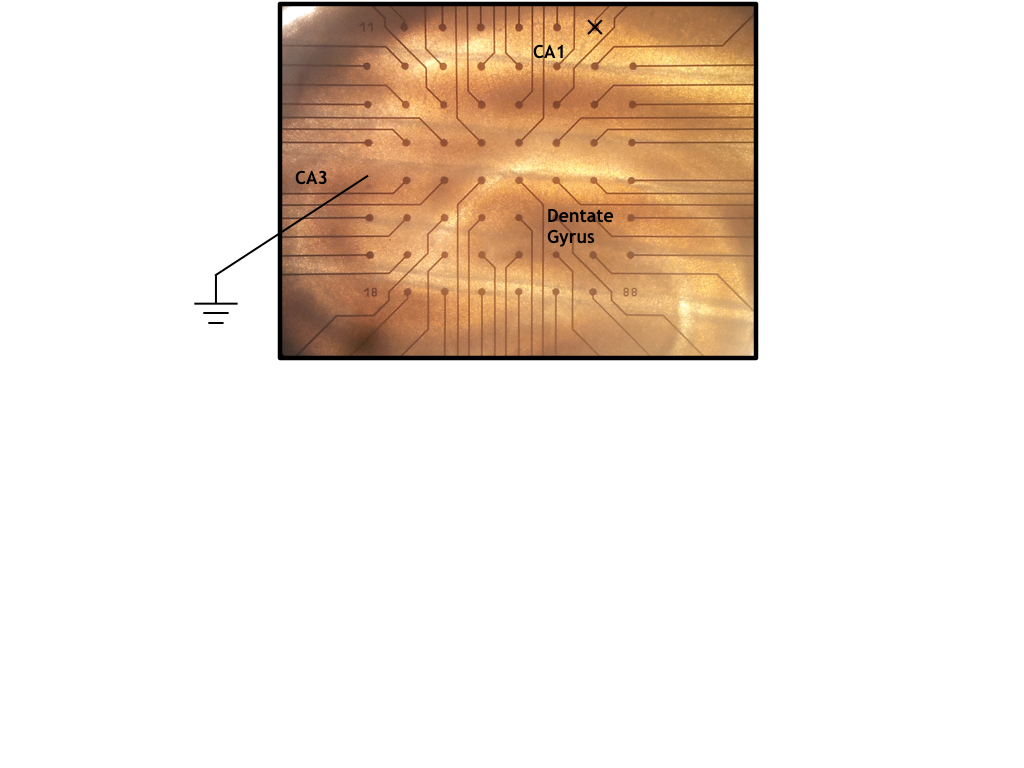}
\caption{\small \newline 
Microscope image of a hippocampal slice on the multi-electrode array. The inter-electrode distance is 200 $\mathrm{\mu m}$. The reference electrode (just below the middle of the leftmost column) and the unused electrode (just to the left of the upper right corner) are indicated. We also give the approximate positions of the regions CA1, CA3, and the Dentate Gyrus as can be determined by usual anatomical considerations.
\label{MEA}} 
\end{figure}

Our dataset was acquired with a multi-electrode array system from Multi Channel Systems GmbH comprising 60 titanium/titanium nitride electrodes of 30 $\mathrm{\mu m}$ diameter arranged in a square grid pattern with 200 $\mathrm{\mu m}$ spacing on a non-conducting glass support. One of the electrodes served as a reference, and another one was not used, leaving 58 active electrodes. The voltage resolution, reflecting the binary representation of the data, was $2^{-16} \times 10 \, \mathrm{mV} \simeq 0.15 \, \mathrm{\mu V}$. An 0.3 mm thick acute hippocampal slice from a 44 day old rat was fixed to the array field with a platina-nylon grid.  See figure \ref{MEA} for a microscope image showing the positions of the electrodes and some of the relevant anatomical structures. Perfusion with a defined artificial cerebrospinal fluid (aCSF) provided the slice with glucose, a physiological salt concentration and osmolality. The layer of fluid above the slice had a thickness of several mm. The electrode potentials were sampled at 25 kHz during 600 s, yielding a total dataset of $870 \times 10^6$ voltage measurements. 

\begin{figure}[h!]
\centering
\includegraphics[width = \linewidth, trim = 0mm 150mm 0mm 0mm, clip]{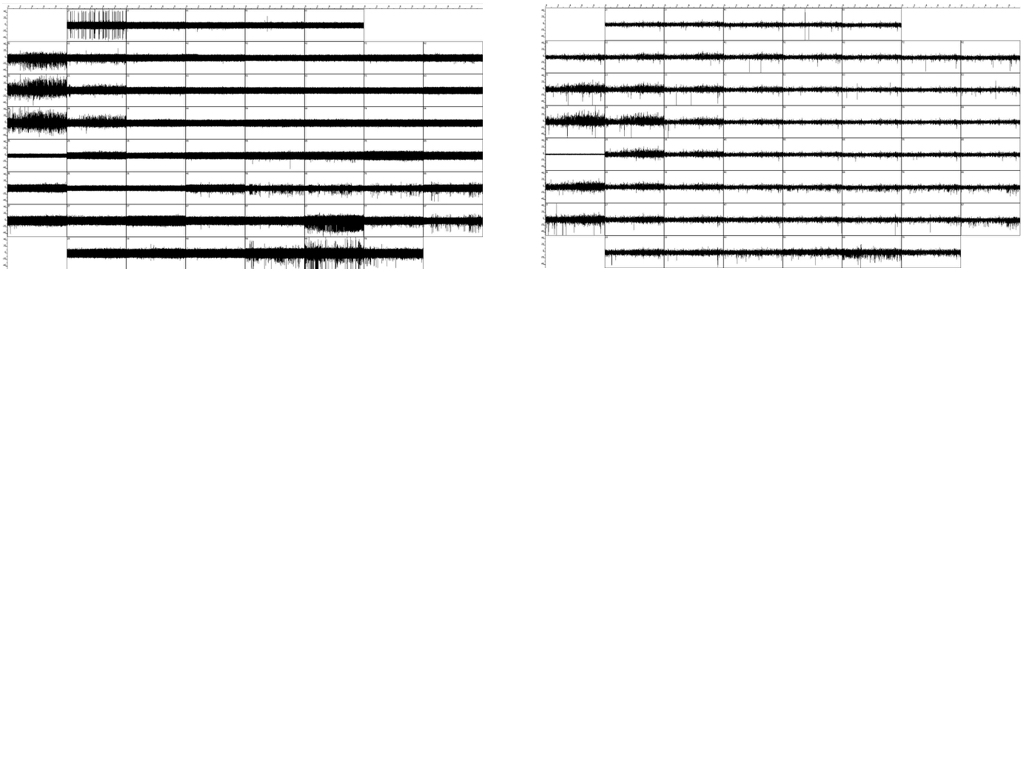}
\caption{\small \newline
{\it Left}: The high-pass ($> 50$ Hz) filtered signals on the 59 electrodes (including the reference electrode). \newline
{\it Right}: The low-pass ($< 50$ Hz) filtered signals. \newline
These figures are produced with a 2nd order Butterworth IIR filter. Each square comprises the entire 10 min registration and a $\pm 50 \, \mathrm{\mu V}$ voltage interval. 
\label{filter}} 
\end{figure}

Conventionally, various filters are applied to the measured signals, but several studies demonstrate that such procedures do not remove spike components in local field potentials (subthreshould activity) \cite{Quilichini-Sirota-Buzsaki}\cite{Ray-Hsiao-Crone-Franaszczuk-Niebur}\cite{Ray-Maunsell}, and we will not use this approach. See however figure \ref{filter} for the high and low pass filtered (above or below 50 Hz respectively) raw data recorded on the 59 electrodes (including the reference electrode just below the middle of the leftmost column).  

\newpage

\subsection{Spikes}
Since the spikes are not our primary interest, but rather obscure the analysis of the much smaller subthreshold fluctuations, they must be detected and removed from the dataset. We do this by a rather simple algoritm, which certainly leaves much room for improvements but is sufficient for our purposes: To detect spike events on an electrode at spatial point $\br$, we consider the difference $d (\br, t)$ between the potential $p (\br, t)$ at time $t$ and its average during a preceeding time interval of some length $\Delta t_\mathrm{average}$. We consider a spike to be fired at time $t$ if the magnitude $| d (\br, t) |$ of this deviation then attains its maximum in the time window of length $2 \Delta t_\mathrm{window}$ centered at $t$ and exceeds a threshold value $d_\mathrm{threshold}$. In view of the typical ms timescale of the action potential dynamics, we used $\Delta t_\mathrm{average} = 10$ ms and $\Delta t_\mathrm{window} = 2$ ms. Concerning the threshold value, the value $d_\mathrm{threshold} = 20 \, \mathrm{\mu V}$ certainly misses many true but smaller spike events, but since our goal here is merely to remove large events that would interfere with the subsequent analysis, this is not a matter of great concern to us. On the other hand, picking a too low threshold value would give many false positives, and would lead us to remove large time intervals of intense neural activity, i.e. precisely the data that is our prime interest. In any case, the precise values of these parameters are not critical for our discussion.

A spike at time $t$ can now be removed by replacing the true potential $p (\br, t^\prime)$ in the interval $t - \Delta t_\mathrm{spike} < t^\prime < t + \Delta t_\mathrm{spike}$ for some time $\Delta t_\mathrm{spike}$ by the linear interpolating function 
\beq
p_\mathrm{linear} (\br, t^\prime) = \frac{t^\prime - t + \Delta t_\mathrm{spike}}{2 \Delta t_\mathrm{spike}} p (\br, t + \Delta t_\mathrm{spike})- \frac{t^\prime - t - \Delta t_\mathrm{spike}}{2 \Delta t_\mathrm{spike}} p (\br, t - \Delta t_\mathrm{spike}) .
\eeq
We used $\Delta t_\mathrm{spike} = 2$ ms, which in view of the observed spiking frequency leads to an almost negligible loss of subthreshold data, while still cutting out all large potential deviations. Henceforth, $p (\br, t)$ will always refer to the potential with all the spikes removed in this way. 

\subsection{The stochastic field theory}
Once the spikes have been removed, our aim is to describe the dynamics of the remaining subthreshold fluctuations. Our approach is to construct a simple model of this as a stochastic process which reproduces the main features of our dataset. Viewing the potential as the sum of a very large number of independent small contributions from different sources indicates (by the central limit theorem of statistics) that it should be normally distributed. This agrees well with the properties of our dataset, and it is thus a reasonable first approximation to limit ourselves to Gaussian models. We choose the reference potential so that the expectation value of the potential $p (\br, t)$ vanishes at all spatial points $\br$ and times $t$:
\beq
\langle p (\br, t) \rangle = 0 .
\eeq
All information is now contained in the two-point function $\langle p (\br_1, t_1) p (\br_2, t_2) \rangle$, and the higher-point functions can be expressed in terms of this by the Isserlis' theorem (in statistical physics mostly known as Wick's theorem), e.g.
\beq
\langle p (\br_1, t_1) p (\br_2, t_2)  p (\br_3, t_3) \rangle = 0
\eeq
and
\bea
\langle p (\br_1, t_1) p (\br_2, t_2)  p (\br_3, t_3) p (\br_4, t_4) \rangle & = & \;\;\; \langle p (\br_1, t_1) p (\br_2, t_2) \rangle \langle p (\br_3, t_3) p (\br_4, t_4) \rangle \cr
&& + \langle p (\br_1, t_1) p (\br_3, t_3) \rangle \langle p (\br_2, t_2) p (\br_4, t_4) \rangle \cr
&& + \langle p (\br_1, t_1) p (\br_4, t_4) \rangle \langle p (\br_2, t_2) p (\br_3, t_3) \rangle . \cr
& & 
\eea
Clearly, average neural activity depends both on the spatial location (related to different anatomical structures) and on time (reflecting the appearance of specific events during the course of the registration). However, in particular in view of the finite amount of data available, a natural first step of the analysis is to disregard these aspects. To begin with, we will thus make the assumption that the stochastic process is stationary in time as well as homogeneous and isotropic in space. We then have
\beq \label{covariance_function_definition}
\langle p (\br_1, t_1) p (\br_2, t_2) \rangle = S (| \br_2 - \br_1 |, | t_2 - t_2 |) 
\eeq
for some covariance function $S (\rho, \tau)$ which will be our primary object of study. Both of these assumptions certainly represent important oversimplifications, and later in the paper we will consider more general models.

On short  time-scales (up to about 100 ms or so), the potential $p (\br, t)$ fluctuates around a slowly varying equilibrium potential $\mu (t)$ that is more or less independent of the spatial position $\br$. We propose to describe this by a local, Gaussian, Markovian stochastic model of the form 
\beq \label{Markov}
\frac{\partial p (\br, t)}{\partial t} = - \gamma \left(p (\br ,t)  - \mu (t) \right) + \alpha \nabla^2 p (\br, t) + \xi (\br ,t) .
\eeq
Here $\nabla^2 = \sum_{i = 1}^2 \partial_i \partial_i$ is the Laplacian operator in two spatial dimensions. The relaxation constant $\gamma$ represents the tendency of the potential to return to its equilibrium value $\mu (t)$, and the diffusion constant $\alpha$ represents the tendency of spatial inhomogeneities to be smoothed out. The last term, which in stochastic modelling is usually referred to as `noise', represents the contributions from the neural activity of a large number of neurons in the vicinity, much as molecular impacts drive Brownian motion. The usefulness of this description is related to the time scale of changes in the equilibrium potential $\mu (t)$ being larger than about 100 ms. For more background on statistical field theory, see e.g. \cite{Itzykson-Drouffe}.

With initial data given in the far past so that its influence can be neglected, the solution to this equation is
\bea \label{Green_solution}
p (\br, t) = \int_{- \infty}^t d t^\prime \int d^2 \br^\prime G (\br - \br^\prime, t - t^\prime) \left( \gamma \mu (t^\prime) + \xi (\br^\prime, t^\prime) \right) ,
\eea
where the Green's function
\beq
G (\br - \br^\prime, t - t^\prime) = \frac{\exp \left(- \gamma (t - t^\prime) - \frac{(\br - \br^\prime)^2}{4 \alpha (t - t^\prime)} \right)}{4 \pi \alpha (t - t^\prime)}
\eeq
obeys the differential equation
\beq
\frac{\partial}{\partial t} G (\br - \br^\prime, t - t^\prime) = \left( - \gamma + \alpha \nabla^2 \right) G (\br - \br^\prime, t - t^\prime)
\eeq
and the initial condition
\beq
G (\br - \br^\prime, 0) = \delta^{(2)} (\br - \br^\prime) .
\eeq
Here and in the sequel, spatial integrals $\int d^2 \br$ are always taken over the infinitely extended plane. Boundary conditions at infinity (provided by the decay of the Green's function) are such that partial integrations do not generate any boundary contributions. The idea of equation \ref{Green_solution} and similiar equations below is that because of the linearity of the model (\ref{Markov}), there is a linear relationship between the driving input (represented by the last factor of the integrand) and the potential. The properties of the Green function ensure that this is indeed a solution to equation \ref{Markov}. (For a further discussion on Green's function techniques for solving linear partial differential equations, see e.g. \cite{Arfken-Weber-Harris}.)

The equilibrium potential $\mu (t)$ and the driving term $\xi (\br, t)$ are both assumed to have vanishing expectations values 
\bea
\left \langle \mu (t) \right \rangle & = & 0 \cr
\left \langle \xi (\br, t) \right \rangle & = & 0 
\eea
leading indeed to a vanishing expectation value for the potential $p (\br, t)$. We furthermore assume the covariance function of $\mu (t)$ to be given by some slowly varying function $S_{\mu^2} (\tau)$, whereas the driving term is assumed to be white both in space and time and uncorrelated with $\mu (t)$:
\bea
\langle \mu (t) \mu (t^\prime) \rangle & = & S_{\mu^2} (| t - t^\prime |) \cr
\left \langle \xi (\br, t) \xi (\br^\prime, t^\prime) \right \rangle & = &  \sigma^2 \delta^{(2)} (\br - \br^\prime) \delta (t - t^\prime) \cr
\left \langle \xi (\br, t) \mu (t^\prime) \right \rangle & = & 0  .
\eea
Here the constant $\sigma^2$ represents the intensity of the neural activity. Accordingly, we can now decompose the covariance function appearing in (\ref{covariance_function_definition}) as
\beq \label{slow_fast}
S (\rho, \tau) = S_\mathrm{slow} (\tau) + S_\mathrm{fast} (\rho, \tau) .
\eeq
 
The first term in (\ref{slow_fast}) represents the contributions from the slow oscillations and can be expressed in terms of the covariance function $S_{\mu^2} (\tau)$ of the equilibrium potential. More precisely
\bea
S_\mathrm{slow} (\tau) & = & \gamma^2 \int_{-\infty}^0 d t^\prime \int d^2 \br^\prime \int_{-\infty}^\tau d t^{\prime \prime} \int d^2 \br^{\prime \prime} \cr
& & G (- \br^\prime, - t^\prime) G (- \br^{\prime \prime}, \tau - t^{\prime \prime}) S_{\mu^2} (| t^\prime - t^{\prime \prime} |) \cr
\cr
& = & \gamma^2 \int_{-\infty}^0 d t^\prime \int_{-\infty}^\tau d t^{\prime \prime} \exp \left(- \gamma (\tau -t^\prime - t^{\prime \prime}) \right) S_{\mu^2} (| t^\prime - t^{\prime \prime} |) \cr 
\cr
& = & \gamma \exp (- \gamma \tau) \int_0^\tau d \tau^\prime \cosh (- \gamma \tau^\prime) S_{\mu^2} (\tau^\prime) \cr
& & + \gamma \cosh (- \gamma \tau) \int_\tau^\infty d \tau^\prime \exp (- \gamma \tau^\prime) S_{\mu^2} (\tau^\prime) .
\eea
In principle, this may be inverted to express $S_{\mu^2} (\tau)$ in terms of $S_\mathrm{slow} (\tau)$:
\beq
S_{\mu^2} (\tau) = \left( 1 - \frac{1}{\gamma^2} \frac{\partial^2}{\partial \tau^2} \right) S_\mathrm{slow} (\tau) .
\eeq
Because of the second derivative, it is however difficult to achieve an accurate estimate of $S_{\mu^2} (\tau)$ with the available data, and we will not develop this approach further. 

The second term in (\ref{slow_fast}) represents the contributions from the driving term and can be expressed in terms of the intensity $\sigma^2$. A short computation gives
\bea \label{Sfast}
S_\mathrm{fast} (\rho, \tau) & = & \int_{-\infty}^0 d t^\prime \int d^2 \br^\prime \int_{-\infty}^\tau d t^{\prime \prime} \int d^2 \br^{\prime \prime} \cr
& & G (- \br^\prime, - t^\prime) G ( {\bf \rho} - \br^{\prime \prime}, \tau - t^{\prime \prime}) \sigma^2 \delta^{(2)} (\br^\prime - \br^{\prime \prime}) \delta (t^\prime - t^{\prime \prime}) \cr
\cr
& = & \sigma^2 \int_{- \infty}^0 d t^\prime \int d^2 \br^\prime \cr
& & \frac{\exp \left(- \gamma (\tau - 2 t^\prime) - \frac{(- \br^\prime)^2}{4 \alpha (- t^\prime)} - \frac{({\bf \rho} - \br^\prime)^2}{4 \alpha (\tau - t^\prime)} \right)}{(4 \pi \alpha)^2 (- t^\prime) (\tau - t^\prime)} \cr
\cr
& = & \sigma^2 \int_{- \infty}^0 d t^\prime \int d^2 \br^\prime \cr
& & \frac{\exp \left(- \gamma (\tau - 2 t^\prime) - \frac{\tau - 2 t^\prime}{4 \alpha (- t^\prime) (\tau - t^\prime)} \left(\br^\prime - \frac{(- t^\prime)  \rho}{\tau - 2 t^\prime} \right)^2 - \frac{\rho^2}{4 \alpha (\tau - 2 t^\prime)} \right)}{(4 \pi \alpha)^2 (- t^\prime) (\tau - t^\prime)} \cr
\cr
& = & \sigma^2 \int_{- \infty}^0 d t^\prime \frac{\exp \left( - \gamma (\tau - 2 t^\prime) - \frac{\rho^2}{4 \alpha (\tau - 2 t^\prime)} \right)}{4 \pi \alpha ( \tau - 2 t^\prime)} . 
\eea
For fixed $\rho$ or $\tau$, this is a monotonously decreasing function of $\tau$ or $\rho$ respectively. In general it cannot be expressed in terms of any well known elementary or special functions. However, for vanishing spatial separation, i.e. $\rho = 0$, it is given by
\bea \label{equal_space}
S_\mathrm{fast} (0, \tau) & = & \sigma^2 \int_{- \infty}^0 d t^\prime \frac{\exp \left( - \gamma (\tau - 2 t^\prime) \right)}{4 \pi \alpha ( \tau - 2 t^\prime)} \cr
\cr
& = & \frac{\sigma^2}{8 \pi \alpha} \Gamma (0, \gamma \tau) \cr
\cr
& = & \frac{\sigma^2}{8 \pi \alpha} \left( - \log (\gamma \tau) - \gamma_\mathrm{EM} + {\cal O} (\gamma \tau) \right) ,
\eea
where $\Gamma$ is the (upper) incomplete Gamma-function and $\gamma_\mathrm{EM} = 0.5772 \ldots$ is the Euler-Mascheroni constant. Similarly, for vanishing temporal separation, i.e. $\tau = 0$, we instead have
\bea \label{equal_time}
S_\mathrm{fast} (\rho, 0) & = & \sigma^2 \int_{- \infty}^0 d t^\prime \frac{\exp \left( - \gamma (- 2 t^\prime) - \frac{\rho^2}{4 \alpha (- 2 t^\prime)} \right)}{4 \pi \alpha (- 2 t^\prime)} \cr
\cr
& = & \frac{\sigma^2}{4 \pi \alpha} K_0 \left( \sqrt{\gamma / \alpha} \, \rho \right) \cr
\cr
& = & \frac{\sigma^2}{4 \pi \alpha} \left( - \log \left( \sqrt{\gamma / \alpha} \, \rho \right) - \gamma_\mathrm{EM} + \log 2  + {\cal O} \left( \sqrt{\gamma / \alpha} \, \rho \right) \right) ,
\eea
where $K_0$ is a modified Bessel-function.

The most important aspects of the results (\ref{equal_space}) and (\ref{equal_time}) are that they exhibit the logarithmic dependence of the covariance function for short temporal and spatial separations respectively with coefficients that are directly related to the parameters of the model. Such logarithmic behaviour is a generic feature of field theories in two spatial dimensions regardless of the details of the model, but does not hold in other dimensions.

\section{Results}
\subsection{Spikes}
\begin{figure}[h!] 
\centering
\includegraphics[width = \linewidth, trim = 0mm 120mm 0mm 0mm, clip]{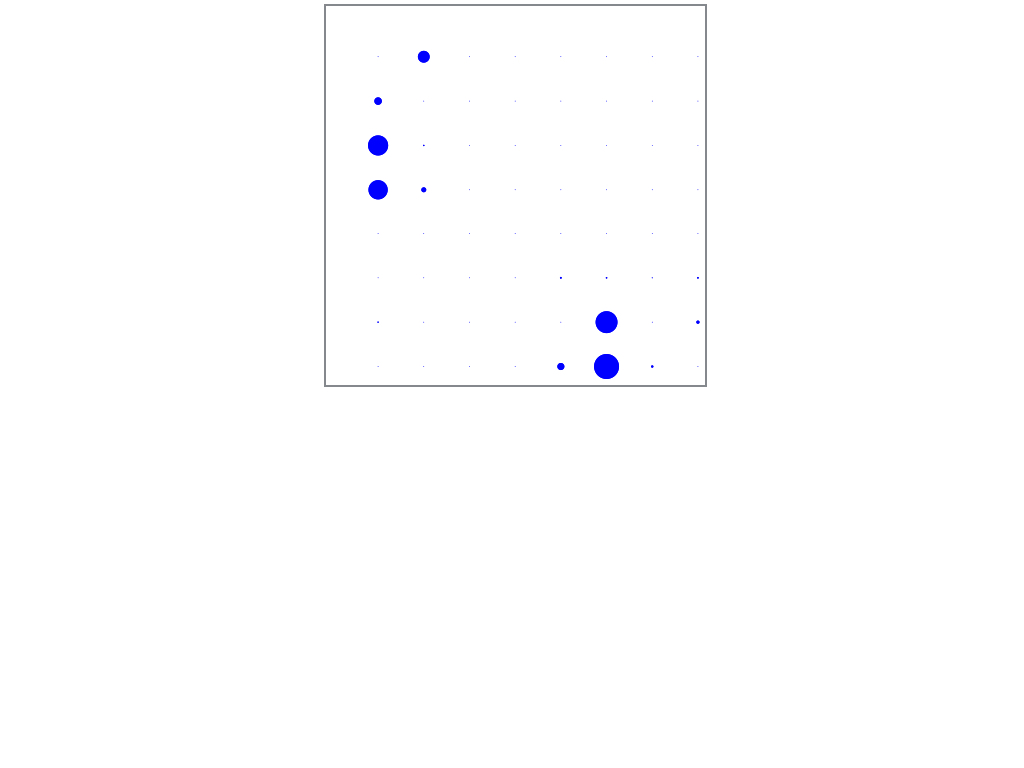}
\caption{\small \newline
Individual spiking frequencies detected on the different electrodes represented by the radii of the dots. The most spiking electrode (in the Dentate gyrus) had a spiking frequency of about 1.7 Hz. \newline
\label{spatial_spikes}} 
\end{figure}

\begin{figure}[h!] 
\centering
\includegraphics[width = \linewidth, trim = 0mm 150mm 0mm 0mm, clip]{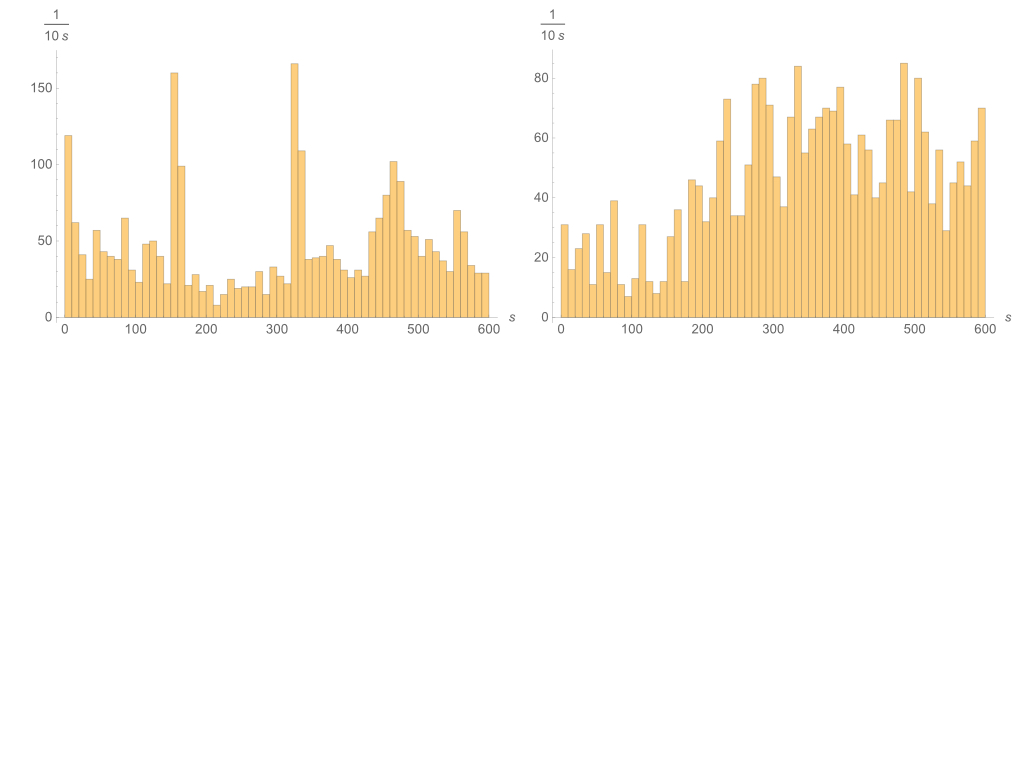}
\caption{\small \newline
{\it Left}: Histogram of the temporal distribution of spikes during the 600 s registration in the Dentate Gyrus in 10 s bins. \newline
{\it Right}: Histogram of the temporal distribution of spikes during the 600 s registration in the CA3 region in 10 s bins. \newline
\label{temporal_spikes}}
\end{figure}

\begin{figure}[h!] 
\centering
\includegraphics[width = \linewidth, trim = 0mm 180mm 0mm 0mm, clip]{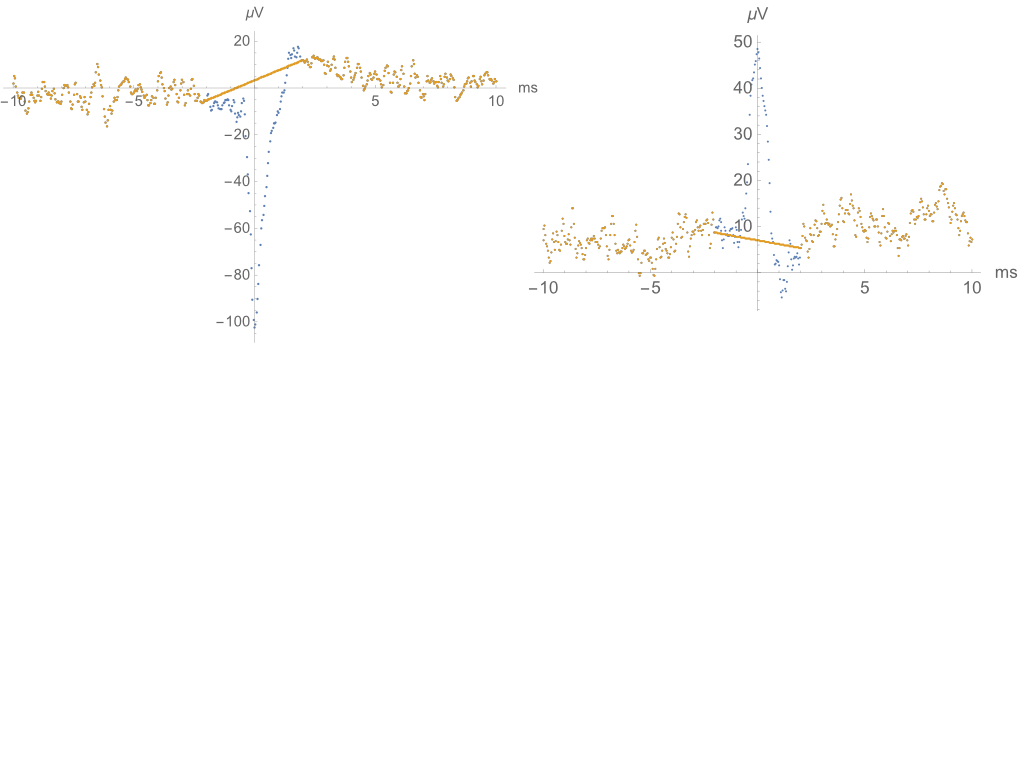}
\caption{\small \newline
{\it Left}: Example of a negative spike (potential as a function of time) in the Dentate Gyrus and its removal by linear interpolation. \newline
{\it Right}: Example of a positive spike in the CA3 region and its removal by linear interpolation. \newline
\label{wave_forms}} 
\end{figure}

With our choices $\Delta t_\mathrm{average} = 10$ ms, $\Delta t_\mathrm{refractory} = 2$ ms, and $d_\mathrm{threshold} = 20 \, \mathrm{\mu V}$, the dataset had a total spike firing frequency of
\beq
\nu_\mathrm{total} \simeq 9.5 \, \mathrm{Hz} .
\eeq
The spikes were rather unevenly distributed, both in time over the 600 s registration and over the 58 electrodes: About 49\% of all spikes were fired in the Dentate Gyrus, where they were mostly negative and tended to occur in short burst of less than 1 s, and 48\% were fired the CA3 region, where they were mostly positive and the spiking frequency fluctuated on times scales of about 100 s. (The remaining 3\% tended to occur in the DG/CA-3 intermediate area.) Although the total number of spikes in these two areas were very nearly equal, their temporal distributions were quite different and give no evidence for any causal connection. See figures \ref{spatial_spikes} and \ref{temporal_spikes}  for the spatial and temporal distribution of the spikes. See figure \ref{wave_forms} for examples of negative and positive spikes and their removal by linear interpolation.

\subsection{Artefacts}
\begin{figure}[h!] 
\centering
\includegraphics[width = \linewidth, trim = 0mm 160mm 0mm 0mm, clip]{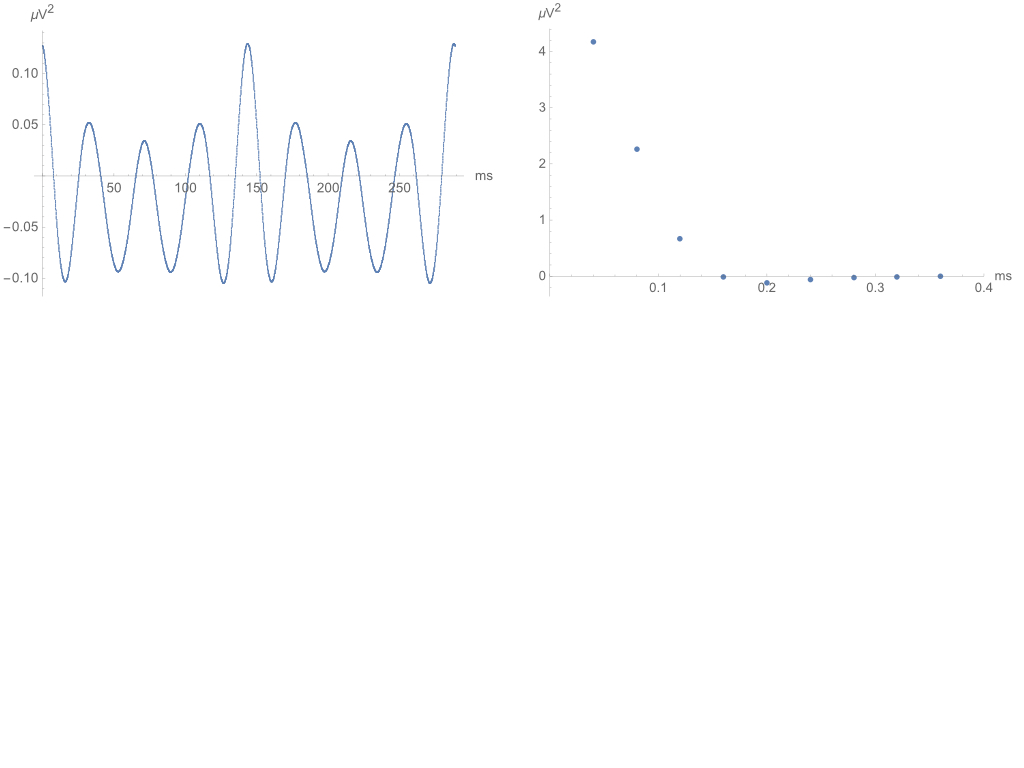}
\caption{\small \newline
{\it Left}: Two periods of the long-term artefact with 145 ms periodicity. This was obtained by subtracting a 145 ms moving average from the raw-data covariance function. For time lags exceeding about 1 s, the difference had an almost perfectly periodic appearance, which was then extend down to zero time lag. \newline
{\it Right}: The rapidly decaying artefact due to errors in the individual voltage measurements. \newline
\label{artefacts}} 
\end{figure}

After the spikes had been removed, we computed the covariance function $S (\rho, \tau)$ by using the entire dataset sampled at 25 kHz. The magnitude of the covariance function $S (0, 0)$ at vanishing spatial and temporal separation, i.e. the variance of the signal, had a magnitude of about 15 $\mathrm{\mu V}^2$. Two features of the covariance function $S (\rho, \tau)$ appeared to be artefacts without biological significance:
\begin{itemize}
\item
There was an almost perfectly periodic component with a period of about 145 ms (corresponding to 6.9 Hz with some overtones) and a maximal amplitude of about 0.12 $\mu V^2$ that persisted essentially undamped until $\tau = 10$ s or more. The extreme and persistent regularity of this phenomenon makes it clear that it originated within the electronics of the multi-electrode array system. Although the magnitude was quite modest, we still found it advantageous (and straightforward) to subtract this component from the covariance function, since its time scale was so close to those of biological relevance.
\item
For $\rho = 0$, i.e. at vanishing spatial separation, there was a component with a pronounced peak in the interval $0 < \tau < 0.2$ ms (i.e. during 5 sampling intervals) with a maximal amplitude of about 4 $\mu V^2$. The short spatial range (less than the electrode spacing) and the short time-scale involved strongly suggested that this phenomenon was due to essentially independent errors in the individual voltage measurements (about 2 $\mu V$) with an extremely short correlation time (about 0.2 ms). Because of the large magnitude, it was necessary to take this component properly taken into account, although its time scale of course was much shorter than those of biological phenomena.
\end{itemize}
See figure \ref{artefacts} for the appearance of these two artefact components in the covariance function. Henceforth $S (\rho, \tau)$ will always refer to the covariance function after these artefacts had been removed by subtracting the two temporal profiles exhibited in the figure from the raw-data covariance function.

\subsection{Slow dynamics}
\begin{figure}[h!] 
\centering
\includegraphics[width = \linewidth, trim = 0mm 80mm 0mm 0mm, clip]{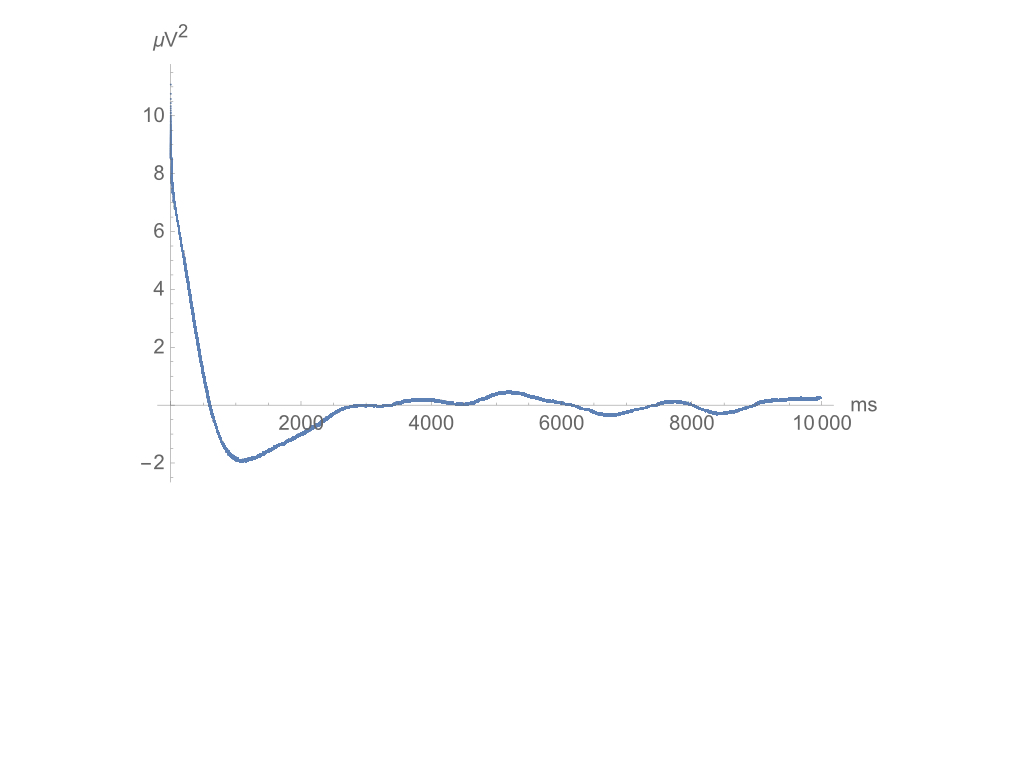}
\caption{\small \newline
Long-time covariance at vanishing spatial separation as a function of time lag. \newline
\label{long_covariance}} 
\end{figure}

For large values of $\tau$  ($> 100$ ms), the main feature of the covariance function $S (\rho, \tau)$  was a damped oscillatory behaviour mainly in the 0 - 2 Hz frequency range, which was largely independent of the spatial separation $\rho$. This slow oscillation is possibly of biological relevance, but we will not attempt to analyze or model it in the present paper. (See e.g. \cite{Buzsaki-Draguhn} for a review of local field potentials with different frequency bands.) See figure \ref{long_covariance} for this long-time behaviour of the covariance function. We took $S_\mathrm{slow} (\tau)$ in (\ref{slow_fast}) to be given by $S (\rho_\mathrm{large}, \tau)$ for $\rho_\mathrm{large} = 1.7$ mm, i.e. the largest spatial separation available to us. As can be seen from figure 8 (lowest curve in the left panel), for such a large spatial separation the covariance was essentially independent of the time lag up to about 100 ms, so $S_\mathrm{fast} (\rho_\mathrm{large}, \tau)$ is negligible. 

\subsection{Fast dynamics}
\begin{figure}[h!] 
\centering
\includegraphics[width = \linewidth, trim = 0mm 160mm 0mm 0mm, clip]{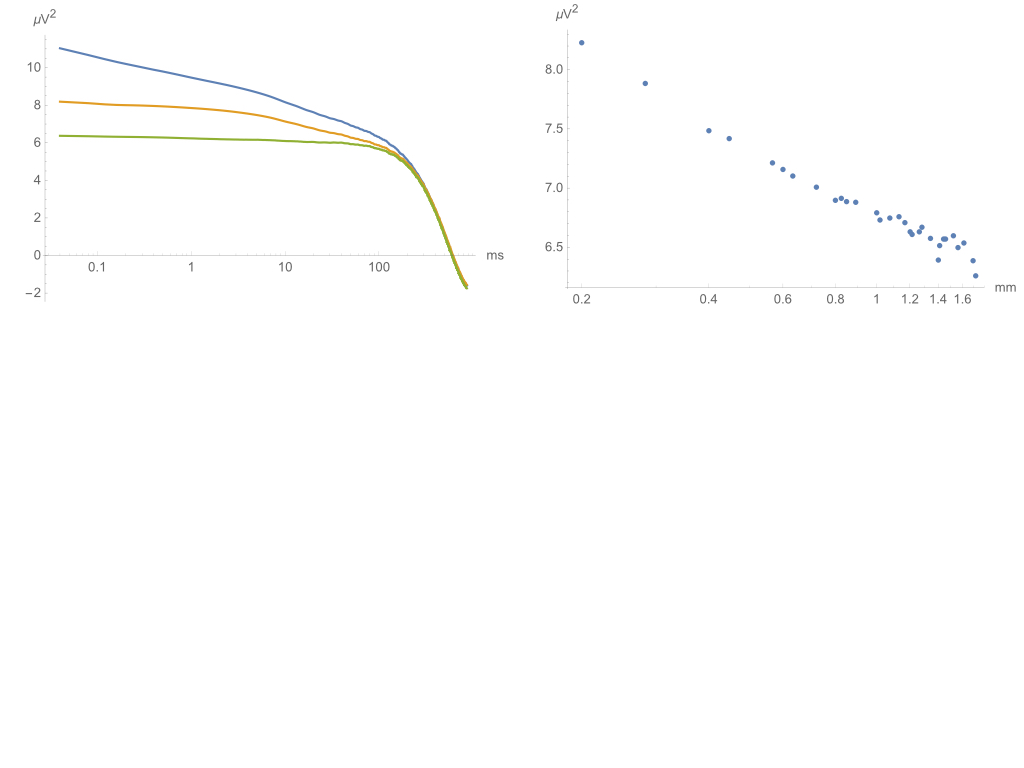}
\caption{\small \newline
{\it Left}: Short-time covariance at spatial separations 0, 0.2 and 1.7 mm (top, middle and lower curve) as a function of time lag. \newline 
{\it Right}: Covariance at vanishing time lag as a function of spatial separation. \newline
In all cases, we exhibit an average over all pairs of electrodes with the indicated spatial separation. 
\label{short_covariance}} 
\end{figure}

For small values of $\tau$ ($< 100$ ms), the covariance function $S (0,\tau)$ at vanishing spatial separation indeed increased logarithmically as $\tau$ approaches zero. For $\rho > 0$, this increase was cut off so that the equal time covariance $S (\rho, 0)$ has a finite value that increases logarithmically as $\rho$ approaches zero. See figure \ref{short_covariance} for the temporal and spatial dependence of the short-time covariance function. Note the logarithmic abscissa axis in these figures!

\begin{figure}[h!] 
\centering
\includegraphics[width = \linewidth, trim = 0mm 150mm 0mm 0mm, clip]{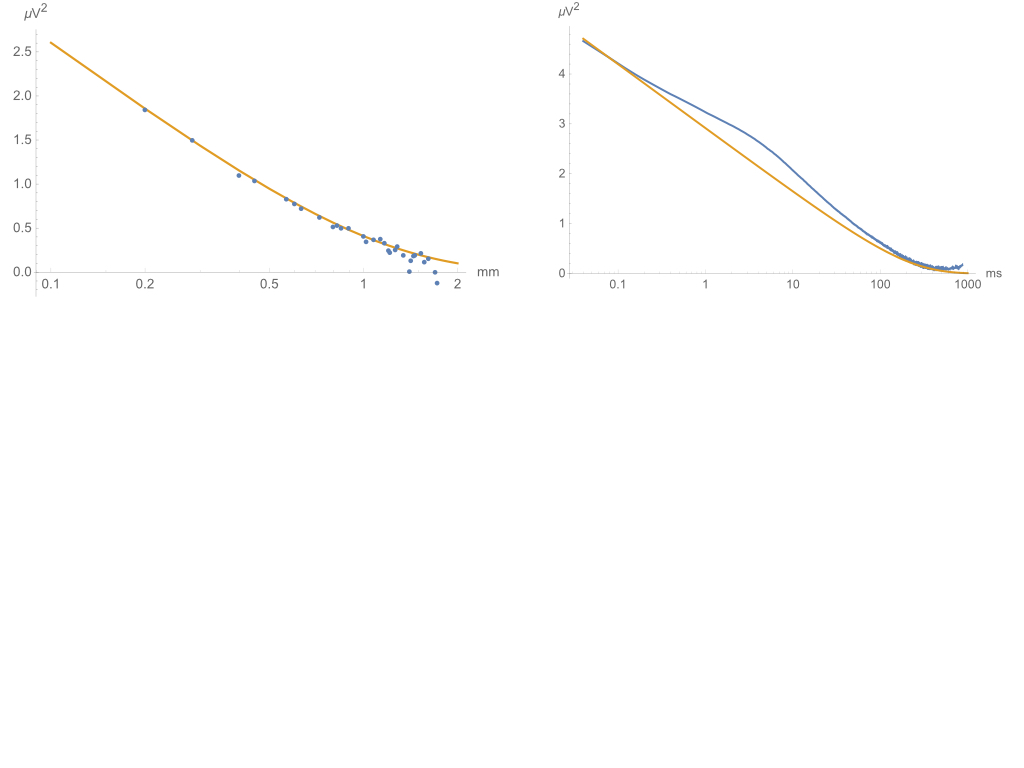}
\caption{\small \newline
{\it Left}: Fitting $S_\mathrm{fast} (\rho, 0)$ to equation \ref{equal_time}. (Equal time covariance as a function of spatial separation.)\newline
{\it Right}: Fitting $S_\mathrm{fast} (0, \tau)$ to  equation \ref{equal_space}. (Covariance as a function of time lag at vanishing spatial separation.)\newline
\label{fit}} 
\end{figure}

The measured values of $S_\mathrm{fast} (\rho, \tau) = S (\rho, \tau) - S (\rho_\mathrm{large}, \tau)$ were fitted to the the theoretical prediction (\ref{Sfast}). This is shown in figure \ref{fit} with the parameter values
\bea
\alpha & \simeq & 0.0025 \, \mathrm{mm}^2 \, \mathrm{ms}^{-1} \cr
\gamma & \simeq & 0.0030 \, \mathrm{ms}^{-1} \cr
\sigma^2 & \simeq & 0.035 \, \mathrm{\mu V}^2 \mathrm{mm}^2 \, \mathrm{ms}^{-1} .
\eea
As can be seen from the figure, the agreement between theory and experiment was excellent, providing a convincing and a priori falsifiable confirmation of the validity of our approach and simplifying assumptions. (In view of our still rather restricted dataset, we refrain from quoting any specific uncertainty range of these parameters.) 

The definition and determination of the three quantities $\alpha$, $\gamma$ and $\sigma^2$ constitute the main results of the present work. An equivalent, but in many respects more illuminating presentation of the results is to combine these parameters into characteristic time, length and voltage scales:
\bea
1 / \gamma & \simeq & 330 \, \mathrm{ms} \cr
\sqrt{\alpha / \gamma} & \simeq & 0.91 \, \mathrm{mm} \cr
\sqrt{\sigma^2 / \alpha} & \simeq & 3.7 \, \mathrm{\mu V} .
\eea

The 330 ms time scale of these `fast' fluctuations may seem uncomfortably close to the time scale of the `slow' fluctuations of the equilibrium potential $\mu (t)$ (which seems to be around 1 s). In this context, we remark that the time scale can be generalized to
\beq
T = \frac{1}{\gamma + \alpha (2 \pi / \lambda)^2}
\eeq
for fluctuations of some finite wavelength $\lambda$. In the long wave-length limit $\lambda \rightarrow \infty$ we recover $1 / \gamma$, whereas for the inter-electrode distance $\lambda = 0.2$ mm we instead get $T \simeq 0.4$ ms. So for wave-lengths relevant for investigating the local dynamics, there is no problem with the time scale. We also remark that the 25 kHz sampling frequency is clearly high enough.

The length scale 0.91 mm is comparable to the extent of the entire multi-electrode array. However, the dimensions of the slice of neural tissue are considerably larger, so there is no need to worry about finite size effects. More importantly, the length scale is sufficiently large compared to the $0.2$ mm inter-electrode distance to assure the validity of this experimental approach to the study of subthreshold fluctuations. 

Finally the voltage scale $3.7 \mathrm{\mu V}$ is safely smaller than the spikes (which we have cut off at 20 $\mathrm{\mu V}$). However, it is quite comparable both to the errors in the individual voltage measurements (about 2 $\mathrm{\mu V}$) and the amplitude of the slow fluctuations of the equilibrium potential $\mu (t)$, so it is important to carefully separate these three phenomena. 

\subsection{Activity}
Sofar we have considered the parameters $\alpha$ and $\gamma$ as well as the activity $\sigma^2$ to be constants. This is reasonable for $\alpha$ and $\gamma$, at least if we view these constants as reflecting only the passive electric transport properties of the intracellular medium and not the propagation of signals along the axons. But the activity should rather be described by a function $\sigma^2 (\br, t)$ of space and time, reflecting the characteristics of the neuronal populations in the different anatomical regions as well as the time course of the neural processing. The value $\sigma^2 \simeq 0.035 \, \mathrm{\mu V}^2 \mathrm{mm}^2 \, \mathrm{ms}^{-1}$ that we have determined should thus be regarded as a spatial and temporal average. 

Retracing the steps leading to (\ref{Sfast}), we find that with a non-constant activity $\sigma^2 (\br, t)$, this expression is no longer valid. However, the leading logarithmic divergence of (\ref{equal_space}), which originates from the short distance behaviour of the model, still holds. Since $S_\mathrm{slow} (t)$ is regular for small $t$, we thus have
\beq \label{timelike}
\langle p (\br, t) p (\br, t + \delta t) \rangle = \frac{\sigma^2 (\br, t)}{8 \pi \alpha} \Bigl( - \log (\gamma \delta t) + {\cal O} (1) \Bigr) .
\eeq
The diffusion constant $\alpha$ is of course already known. The values of $\delta t$ can e.g. be chosen in the interval 0.2 ms to 10 ms. Since we have only a single measurement of the potential $p (\br, t)$ for each value of $\br$ and $t$, we can only estimate such expectation values by averaging over a rather large time interval (at least about 100 ms) around $t$, which limits the temporal resolution of the method. 

\begin{figure}[h!] 
\centering
\includegraphics[width = \linewidth, trim = 0mm 90mm 0mm 0mm, clip]{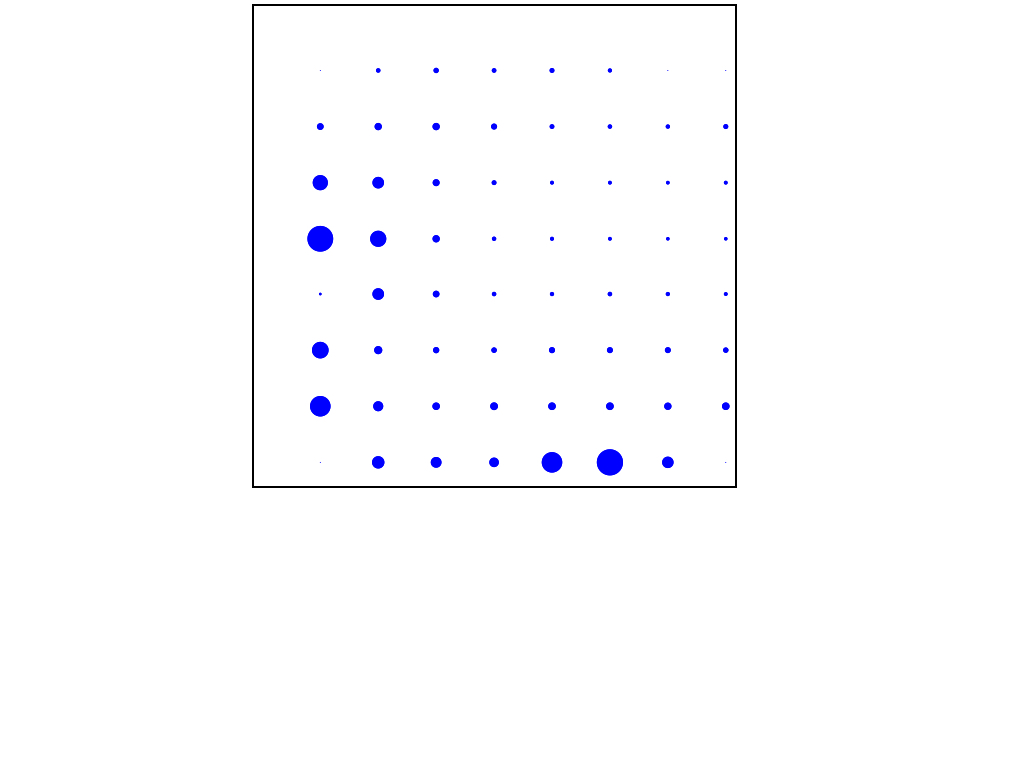}
\caption{\small \newline Temporal mean over the 600 s registration of the activity on the different electrodes. The most active electrode (in the Dentate gyrus) has an activity of about 0.11 $\mathrm{\mu V^2 mm^2 ms^{-1}}$. \newline
\label{spatial_activity}} 
\end{figure}

Averaging over the entire 600 s registration, we found that the temporal mean $\overline{\sigma^2 (\br, t)}$ of the activity was concentrated in the Dentate Gyrus and the CA3 region just like the spikes, but much more spread out. There was however also substantial activity in the area intermediate between these two regions (where essentially no spikes occur), whereas the CA1 region showed very little activity. See figure \ref{spatial_activity} for this spatial distribution of activity. Comparison can be made with the spatial distribution of spikes in figure \ref{spatial_spikes}. 

\begin{figure}[h!] 
\centering
\includegraphics[width = \linewidth, trim = 0mm 30mm 0mm 0mm, clip]{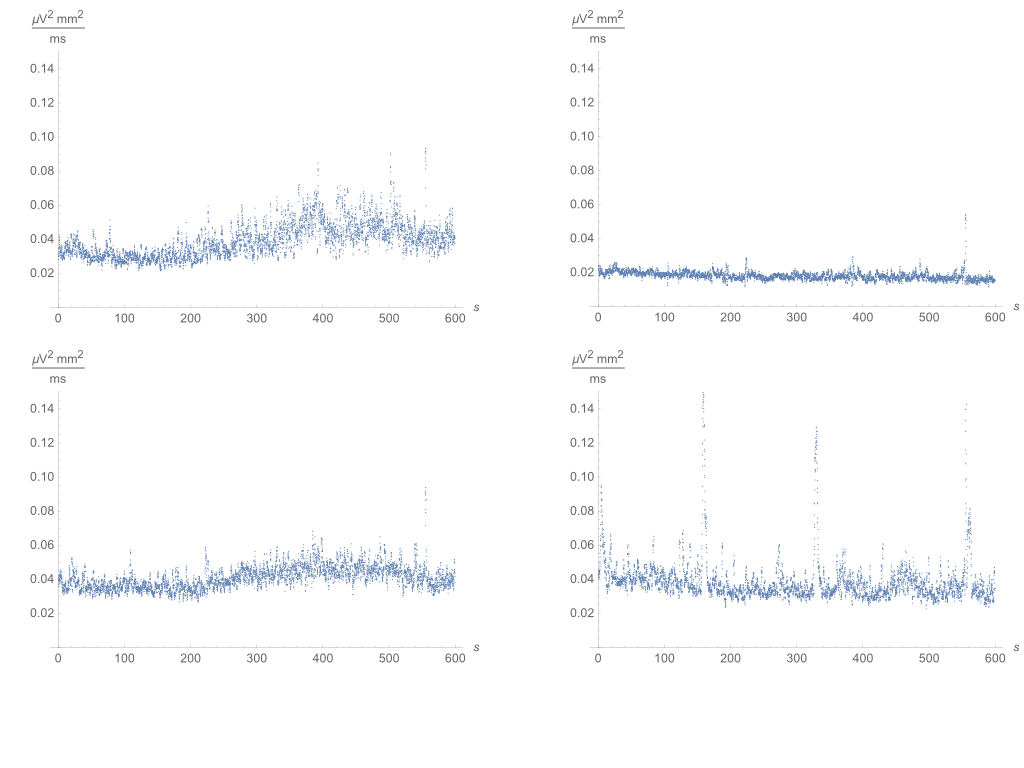}
\caption{\small \newline
Mean activity in each of the four quadrants of the multi-electrode array (roughly corresponding to the Dentate Gyrus, the DG/CA3-intermediate area, the CA3 region and the CA1 region clockwise from the lower right corner) as a function of time during the 600 s registration. The temporal resolution in these graphs is 1 s.
\label{temporal_activity}} 
\end{figure}

Averaging over time intervals of 1 s instead, we could investigate the temporal dependance of the activity in the different regions. We found a clear correlation with the spiking in the Dentate Gyrus and the CA3 region. In the intermediate non-spiking area, the pattern was more reminscent of the CA3 region than the Dentate Gyrus. The CA1 region showed a rather constant lower activity. See figure \ref{temporal_activity}, which should be compared with the corresponding temporal distributions of spikes in figure \ref{temporal_spikes}.

One sees that activity and spiking are indeed different phenomena, although there seems to exist some connection between them. Again, we take the view that the spiking frequency registered on the different electrodes reflects not only what is going at that location in the tissue but also on how a few individual neurons happen to be in more or less close contact with the electrodes.

\subsection{Connectivity}
In contrast to the potential $p (\br, t)$, which is a stochastic variable, we have considered the activity $\sigma^2 (\br, t)$ to be a given function of space and time. Ultimately, one would of course like to formulate some (deterministic or stochastical) dynamical model for it, but we will not pursue this here and instead content ourselves with a purely descriptive treatment. While the activity directly influences the variance of the signal $p (\br, t)$, it shows essentially no correlation with the mean of $p (\br, t)$. This is another indication that the equilibrium potential $\mu (t)$, while serving as a common voltage reference for the entire network, may not be of immediate biological relevance.

A very useful quantity for characterizing the activity $\sigma^2 (\br, t)$ is its covariance function between separate points $\br_1$ and $\br_2$ at some time lag $\Delta t$:
\beq
\mathrm{Cov} \left(\sigma^2(\br_1, t), \sigma^2 (\br_2, t + \Delta t) \right) = \overline{\sigma^2 (\br_1, t) \sigma^2 (\br_2, t + \Delta t)} - \overline{\sigma^2 (\br_1, t)} \; \overline{\sigma^2 (\br_2, t)} .
\eeq
Here and in the sequel, an overline denotes an average over the time $t$. In particular, we have the temporal autocovariance function
\beq
\mathrm{Cov} \left(\sigma^2_\mathrm{mean} (t) , \sigma^2_\mathrm{mean} (t + \Delta t) \right) = \overline{\sigma^2_\mathrm{mean} (t) \sigma^2_\mathrm{mean} (t + \Delta t)} - \overline{\sigma^2_\mathrm{mean} (t)} \; \overline{\sigma^2_\mathrm{mean} (t + \Delta t)} ,
\eeq
where
\beq
\sigma^2_\mathrm{mean} (t) = \frac{1}{\mathrm{Vol}_\Omega} \int_\Omega d^2 \br \, \sigma^2 (\br, t)
\eeq
is the spatial mean of the activity. (We take the domain $\Omega$ to cover the entire multi-electrode array.) Empirically, we find that
\beq \label{temporal_corr}
\mathrm{Cov} \left(\sigma^2_\mathrm{mean} (t) , \sigma^2_\mathrm{mean} (t + \Delta t) \right) \sim \exp \left( - \beta \Delta t \right) ,
\eeq
with decay constant
\beq
\beta \simeq 0.1 \, \mathrm{s}^{-1} .
\eeq
(In these last formulas, the expressions are in fact independent of the time $t$ appearing in the left hand sides.)

\begin{figure}[h!] 
\centering
\includegraphics[width = \linewidth, trim = 0mm 150mm 0mm 0mm, clip]{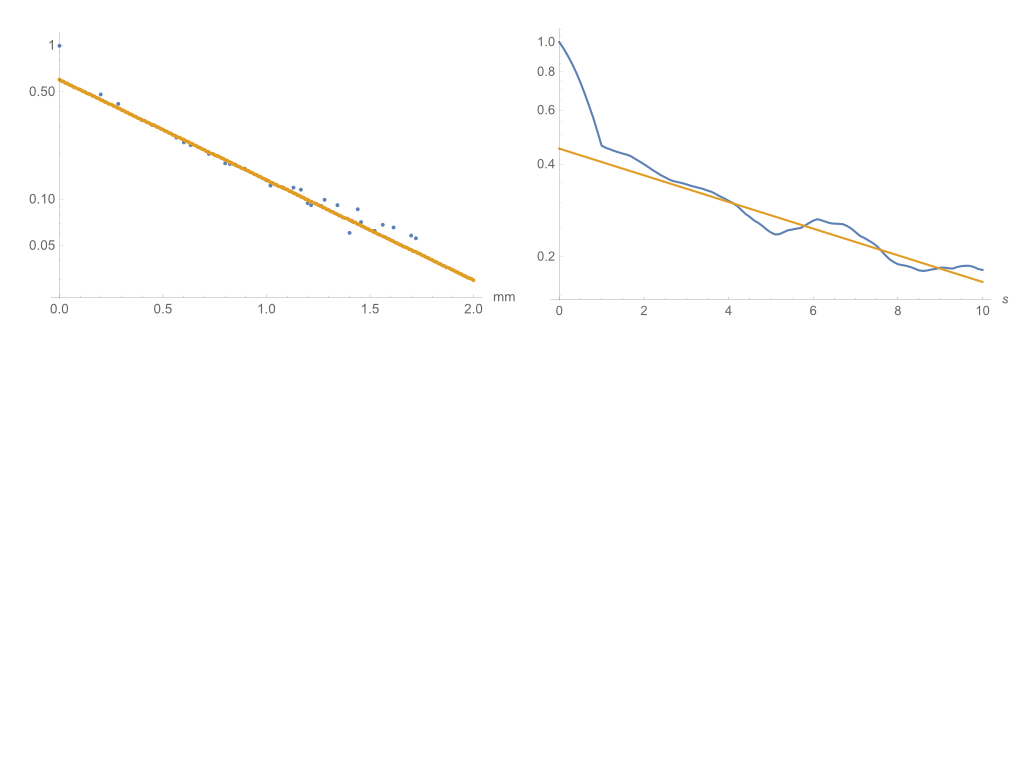}
\caption{\small \newline
{\it Left}: Fitting the spatial autocorrelation function of the activity at equal time to the exponential expression \ref{mean_spatial_corr}. \newline
{\it Right}: Fitting the temporal autocorrelation function of the spatial mean activity to the exponential expression \ref{temporal_corr}. \newline
\label{sigma2-correlation}} 
\end{figure}

Similarly, we can investigate the spatial autocovariance function
\beq \label{spatial_corr}
\mathrm{Cov} \left(\sigma^2 (\br, t) , \sigma^2 (\br + \Delta \br, t) \right) = \overline{\sigma^2 (\br, t) \sigma^2 (\br + \Delta \br, t)} - \overline{\sigma^2 (\br, t)} \; \overline{\sigma^2 (\br + \Delta \br, t)} .
\eeq
Taking the spatial mean, we here find
\beq \label{mean_spatial_corr}
\frac{1}{\mathrm{Vol}_\Omega} \int_\Omega d^2 \br \, \mathrm{Cov} \left(\sigma^2 (\br, t) , \sigma^2 (\br + \Delta \br, t) \right) \sim \exp ( - \kappa | \Delta \br |) ,
\eeq
with decay constant
\beq
\kappa \simeq 1.4 \, \mathrm{mm}^{-1} .
\eeq
See figure \ref{sigma2-correlation} for the corresponding autocorrelation functions (normalized to $1$ for $\Delta t = 0$ and $\Delta \br = 0$ respectively). Note the logarithmic scales! The deviations from exponential decay for small time lags and distances can be attributed to the measurement errors, which in the temporal case are smoothed out over 1 s by our data analysis. 

The behaviour of these covariance functions should be relevant for the understanding of neural connectivity and communication. The exponential decay is qualitatively rather different from the logarithmic behaviour characteristic of passive transport in two spatial dimensions as we have investigated for the potential $p (\br, t)$. This possibly indicates that the activity is propagated by some more `active' mechanism, for which we do not have any specific proposal. However, the temporal scale of about 10 s is long enough that the biological significance of these slow changes may be questioned, in which case they should probably be attributed to drifting conditions during the registration. Indeed, the biologically relevant information transfer in the neural tissue should be encoded in fluctuations of the activity at much shorter time-scales, which we are however unable to probe with our present methods. On the other hand, the spatial scale of about 0.7 mm is quite similar to the scale $\sqrt{\alpha / \gamma} \simeq 0.91 \mathrm{mm}$ set by the diffusion process, and again indicates that the multi-electrode array is adequate for investigating these phenomena.

\begin{figure}[h!] 
\centering
\includegraphics[width = \linewidth, trim = 0mm 150mm 0mm 0mm, clip]{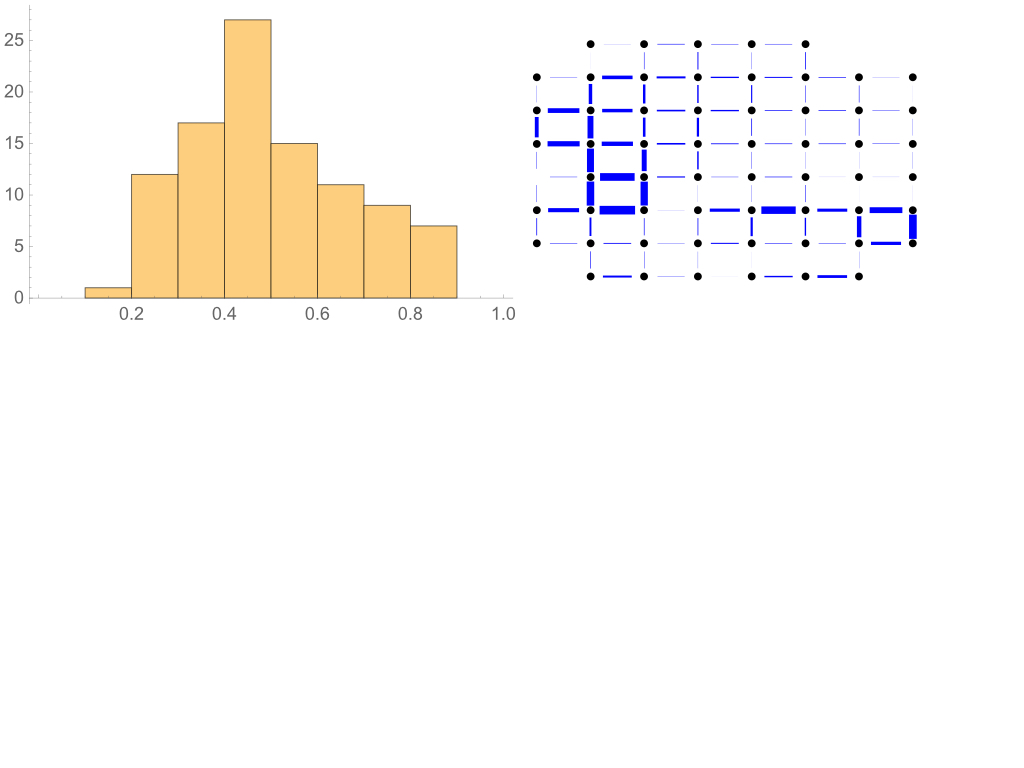}
\caption{\small \newline
{\it Left}: Histogram of the distribution of correlation coefficients for the activity on all pairs of adjacent electrodes. \newline
{\it Right}: The connectivity between adjacent electrodes. The thickness of the lines is proportional to the fourth power of the correlation coefficient for the corresponding pair of electrodes. \newline
\label{connectivity}} 
\end{figure}

Finally, we considered the Pearson correlation coefficient of the activity $\sigma^2(r_1, t)$ and $\sigma^2(r_2, t)$ separately for all pairs of adjacent electrodes, which gives a way of investigating the local neural connectivity. A priori, such a correlation can be weak or strong regardless of the mean and variances of the two activities under consideration. With $| \Delta \br | = 0.2$ mm the average correlation coefficient was about 0.50, but varied considerably between 0.1 and 0.9 for the different pairs. With sufficiently strong inhibitory connections, one could in principle also imagine negative correlation coefficients in the interval -1 to 0, but these did not occur in our dataset. Highly correlated pairs indicated a path of information flow from the Dentate Gyrus to the region CA3 with a hint of a continuation towards CA1, in agreement with the expectations from anatomical considerations. (Actually, our methods cannot determine the direction of this information flow, since the correlation is invariant under the exchange of two electrodes.) See figure \ref{connectivity} for an attempt at a graphical rendering of this connectivity pattern. 

\section{Discussion}
Our main finding is that the local field potential can be remarkably well described by a Gaussian statistical field theory in two space and one time dimension. Depending on the electrical properties of the perfusion liquid above the tissue sample, one may argue that this should be modelled as a three-dimensional rather than a two-dimensional system. This would give a qualitatively rather different model, in which correlations decay as the inverse of the distance rather than logarithmically in both space and time. However, our two-dimensional model fits the data excellently, whereas such a three-dimensional model would be in clear disagreement. So we have been able to make a clear and falsifiable theoretical prediction and verify it experimentally in what we think is a convincing manner. From a perspective of practical electrophysiology, we would like to emphasize that this analysis must be preceeded by a correct elimination of certain artecfacts of no biological significance.

Thus, we have provided a proof of concept of a new approach for studying neural circuit function and applied it to the dataset described above. By our approach, we have described the mean activity (figure \ref{temporal_activity}) and correlation (figure \ref{connectivity}) of subthreshold fluctuations within specific hippocampal sub-regions. It appears that the connection between CA3 and CA1 in this particular isolated hippocampal ex vivo slice preparation is not preserved. Even though this represents a drawback of our used MEA data set, the presented connection between DG and CA3 demonstrates that our approach allows for the identification and visualization of connected sub-regions in isolated brain-slice preparations.

 We also computed specific mean values for the parameters characterizing the duration, spatial distribution and amplitude of subthreshold fluctuations in all hippocampal sub-regions. Since we aim to present a proof-of-concept of our approach by using data sets collected only in only one hippocampal slice preparation, we did not perform a hippocampal sub-region specific classification of subthreshold fluctuations. Of course, such parametric description of sub-neuronal network properties within the hippocampus is quite interesting and will be addressed in future studies.

Our method to extract subthreshold fluctuations out of MEA data sets can be used to uncover spatial and temporal correlations of sub-hippocampal neuronal circuits within brain slice preparations, which can not be achieved by analyzing the localization of spike activity. Indeed, while the possibility to detect spikes is largely determined by the accidental proximity of a neuron to an electrode, the activity as we define it should be a robust concept. Thus, describing the spatial and temporal properties of subthreshold fluctuations in ex vivo or in vitro neuronal circuits may represent a better approach to uncover functional connectivity within neuronal circuits than analysis of synchronous bursting. However, of course also the activity as we have defined it is to a large extent determined by the population of nearby neurons, so it is not obvious to separate these aspects from each other. Indeed, although we have defined the activity without any reference to detected spike events, it still shows a clear correlation with these, and its spatial distribution and correlations agree well with expectations from anatomical considerations. See e.g. \cite{Giugliano} for a discussion of the relationship between collective network phenomena and the spiking of individual neurons.

It would be interesting to try to get a better understanding of the laws underlying the slow dynamics, rather than just describing the resulting equilibrium potential. This can be done with a larger dataset, provided that the longterm stability of the preparation can be assured.

Our model has a small number free parameters, the values of which can be readily determined by fitting the experimentally measured correlations. We expect that these parameters will provide robust and reproducible quantities suitable for comparative studies between brain tissue samples from different anatomical regions and developmental stages under various physiological and patho-physiological conditions. It can also be valuable to study patient-specific neuronal circuits obtained from induced pluripotent stem cell technology. In the future, it would thus be very interesting to apply these methods to more datasets.

In a different direction, the understanding of the passive transport properties of the neural preparation developed in this article should be useful also for analyzing spiking events. Indeed, such an analysis is complicated by the fact that signals spread between the different electrodes, so a natural approach is to begin by reconstructing the local sources of these events by inverse methods based on our model. We plan to return to these issues in forthcoming publications.

\subsubsection*{Acknowledgments}
We have benefitted from discussions with Eric Hanse.\\
The research of M.H. was supported by a grant from the Kristina Stenborg foundation.\\
The research of S.I. was supported by a grant from the Alzheimerfonden (AF-556051).

\end{document}